\documentclass{snmp2003}
\begin{document}
\FirstPageHeading{Vernov}

\ShortArticleName{The Painlev\'e analysis and 
                  construction of special solutions} 

\ArticleName{Construction \ of \ Special \ Solutions \ for \ Nonintegrable
Dynamical \ Systems \ with \ the \ help \ of \ the \ Painlev\'e \ Analysis}

\Author{Sergey Yurievich VERNOV~$^\dag$} 
\AuthorNameForHeading{S.Yu. Vernov} 
\AuthorNameForContents{Vernov S.Yu.}
\ArticleNameForContents{Construction of Special Solutions for Nonintegrable
Dynamical Systems with the help of the Painlev\'e Analysis}

\Address{$^\dag$~Skobeltsyn Institute of Nuclear Physics, Moscow State 
University, Vorob'evy Gory, Moscow, 119992, Russia} 
\EmailD{svernov@theory.sinp.msu.ru} 


\Abstract{The generalized H\'enon--Heiles system
has been considered. In two nonintegrable cases with the help of the
Painlev\'e test new special solutions have been found as Laurent
series, depending on three parameters.
The obtained series converge in some ring.
One of parameters determines the singularity point location,
other parameters determine coefficients
of series.  For some values of these parameters the
obtained Laurent series coincide with the Laurent series of the known
exact solutions.  The Painleve test  can  be  used  not  only to
construct local  solutions as the Laurent series but also to  find
elliptic solutions.}

\section{THE PAINLEV\'E PROPERTY AND INTEGRABILITY}
A Hamiltonian system in a $2s$--dimensional phase space
is called { \it completely integrable } (Liouville integrable) if it
possesses $s$ independent integrals which commute with
respect to the associated Poisson bracket.  When this is the case, the
equations of motion are separable  (at least, in principal) and solutions
can be obtained by the method of quadratures.

When some mechanical problem is studied,
time is assumed to be real, whereas the integrability of motion
equations depends on the behavior of their solutions as functions of
complex time. S.V.~Kovalevskaya\, was the first, who proposed~\cite{Vernov:Kova}
to interpret time as a complex variable and to require that 
the mechanical-problem solutions have to be single-valued functions  
meromorphic in the entire complex plane. 
This idea  led\, S.V.~Kovalevskaya\, to a remarkable result: a
new integrable case (nowadays known as the Kovalevskaya's case) for the
motion of a heavy rigid body about a fixed point was 
discovered~\cite{Vernov:Kova}
(see also~\cite{Vernov:Golubev2,Vernov:Goriely0}).
The Kovalevskaya's result demonstrated that the analytic 
theory of differential equations can be fruitfully applied to mechanical 
and physical problems. The important stage of development of this theory 
was the Painlev\'e classification of ordinary differential equations 
(ODE's) with respect to the types of singularities of their 
solutions~\cite{Vernov:Painleve1}.

Let us formulate the Painlev\'e property for ODE's.
Solutions of a system of ODE's
are regarded as analytic functions, may be
with isolated singularity points~\cite{Vernov:Golubev1, Vernov:Hille}.
A singularity point of a solution is said {\it critical } (as opposed to
{\it noncritical\,}) if the solution is multivalued (single-valued)
in its neighborhood and {\it  movable} if its location
depends on initial conditions.
The {\it general solution} of an ODE of order $N$
is the set of all solutions mentioned in the existence theorem of
Cauchy, i.e. determined by the initial values.
It depends on $N$ arbitrary independent constants.
A {\it special solution} is any solution obtained from the
general solution by giving values to the arbitrary constants.
A {\it singular solution} is any solution which is not special,
i.e. which does not belong to the general solution.

\begin{definition}
 A system of  ODE's has
 \textbf{\textit{ the Painlev\'e
property }} if its general solution has no movable
critical singularity point~\cite{Vernov:Painleve1}.
\end{definition}

Investigations of many dynamical systems~\cite{Vernov:Tabor} show that
systems with the Painlev\'e property are completely integrable. 
Arguments, which clarify the connection between the Painlev\'e
analysis and the existence of motion integrals, are presented
in~\cite{Vernov:ES1,Vernov:ES2}. At the same time 
the integrability of an arbitrary system
with the Painlev\'e property has yet to be proved.  There is not an
algorithm for construction of the additional integral by the Painlev\'e
analysis. It is easy to give an example of an integrable system without
the Painlev\'e property~\cite{Vernov:Kozlov}.  The 
system with the Hamiltonian $H=\frac{1}{2}p^2+f(x)$, 
where $f(x)$ is a polynomial which power is not lower than five, is 
trivially integrable, but its general solution is not a meromorphic 
function. 

 The study of complex-time singularities is a useful tool for the analysis
of not only integrable systems, but also chaotic dynamics~\cite{Vernov:GoTa}.
The Painlev\'e analysis can be connected with the normal form
theory~\cite{Vernov:Goriely}.

The \textbf{\textit{ Painlev\'e test }} is any
algorithm, which checks some necessary
conditions for a differential equation to have the Painlev\'e property.
The original algorithm, developed by P.~Painlev\'e and
used by him to find all the second order ODE's with
the Painlev\'e property, is known as the \ \  \ \ \ \  \ \  \ \ \ \ 
\ \  \ \ \ \  \ \  \ \ \ \ $\alpha$--method.
The method of S.V.~Kovalevskaya is not as general as the
$\alpha$--method, but much more simple.
The remarkable property of this test is that it can be checked in a
finite number of steps. This test can only detect the
occurrence of logarithmic and algebraic branch points.  Up to the present 
there is no general finite algorithmic method to detect the occurrence 
of essential singularities. Different variants
of the Painlev\'e test are compared in~\cite[R.~Conte
paper]{Vernov:Conte0}.

Developing the Kovalevskaya method~\cite{Vernov:Kova} further,
 \  M.J.~Ablowitz, A.~Ramani and H.~Segur
constructed  a new algorithm of the Painlev\'e test for ODE's~\cite{Vernov:ARS}.
 They also were the first to point out the
connection between the nonlinear partial differential equations (PDE's),
which are solvable by the inverse scattering transform method, and ODE's
with the Painlev\'e property. Subsequently the Painlev\'e property for PDE's
was defined and the corresponding Painlev\'e test (the WTC procedure) was
constructed~\cite{Vernov:WTC,Vernov:Weiss4} (see  
also~\cite{Vernov:Conte0,Vernov:Conte2}). With the
help of this test it has been found, that all PDE's, which are solvable by
the inverse scattering transforms, have the Painlev\'e property, may be,
after some change of variables.  For many integrable PDE's, for example,
the Korteweg--de-Vries equation~\cite{Vernov:Tabor}, the
B\"acklund transformations and the Lax representations result from the WTC
procedure~\cite{Vernov:Weiss4, Vernov:Conte1988}. For certain
nonintegrable PDE's special solutions  were constructed using this
algorithm~\cite{Vernov:Tabor2,Vernov:Conte3}.

The algorithm for finding
special solutions for ODE's in the form of finite expansions with respect to
powers of an unknown function has been
constructed in~\cite{Vernov:Weiss1,Vernov:Weiss2}.   This function 
and coefficients have to satisfy some system of ODE's, 
often more simple than an initial one. This method has been used to  
obtain exact special solutions for some nonintegrable
ODE's~\cite{Vernov:GNTZ,Vernov:Sahadevan}.  With the help of the 
perturbative Painlev\'e test~\cite{Vernov:Conte2} a  four-parameter 
generalization of an exact three-parameter solution of the Bianchi IX 
cosmological model has been 
constructed~\cite{Vernov:Bianchi3}.

\section{THE H\'ENON--HEILES HAMILTONIAN}

In the 1960s the models of the star
motion~in an axial-symmetric and  time-independent potentials have been
developed to show either existence or absence
of the third integral for some polynomial potentials.
Due to the symmetry of the potential the considered system
is equivalent to two-dimensional one.
To clarify the question about the existence of the third integral
H\'enon and Heiles~\cite{Vernov:HeHe} considered the behavior of numerically
integrated trajectories. Emphasizing that their
choice does not proceed from experimental data,
they have proposed the Hamiltonian
\[
H=\frac{1}{2}\Big(x_t^2+y^2_t+ x^2+y^2\Big)+x^2y-\frac{1}{3}y^3,
\]
because on the one hand, it is
analytically simple; this makes the numerical computations of trajectories
easy;  on the other hand, it is sufficiently complicated to
 give trajectories which are far from trivial. Indeed, for low energies
the  H\'enon--Heiles system appears to be
integrable, in so much as trajectories (numerically integrated) always lay
on well-defined two-dimensional surfaces. On the other hand,
for high energies many of these integral surfaces are
destroyed, it points on absence of the third integral.

The generalized H\'enon--Heiles system is described by the Hamiltonian:
\begin{equation}
\label{Vernov:1}
H=\frac{1}{2}\Big(x_t^2+y_t^2+\lambda
x^2+y^2\Big)+x^2y-\frac{C}{3}\:y^3
\end{equation}
and the corresponding system of the motion equations:
\begin{equation}
\label{Vernov:2}
\left\{
\begin{array}{l} \displaystyle x_{tt} = -\lambda x -2xy,\\[2.7mm] 
 \displaystyle y_{tt}  
= -y-x^2+Cy^2, \end{array} \right.  
\end{equation} 
where 
$x_{tt}\equiv\frac{d^2x}{dt^2}$ and $y_{tt}\equiv\frac{d^2y}{dt^2}$, \
$\lambda$ and $C$ are numerical parameters.

The generalized H\'enon--Heiles system is a model not only actively 
investigated by various mathematical methods (see~\cite{Vernov:Vernov0203} 
and references therein), but also widely used in 
astronomy and physics, in particular, 
in gravitation~\cite{Vernov:Podolski,Vernov:Kokubun1}.
The models, described by the Hamiltonian (\ref{Vernov:1}) with some additional
nonpolynomial terms, are actively 
studied~\cite{Vernov:Polska,Vernov:Conte4,Vernov:ConteTMF} as well.

Due to the Painlev\'e analysis the following integrable cases 
of~(\ref{Vernov:2}) have been found:
\[ 
\begin{array} {cll} \mbox{(i)} & C=-1,
&\lambda=1,\\ \mbox{(ii)} & C=-6, &\mbox{$\lambda$ is an arbitrary
number},\\ \mbox{(iii)} & C=-16,\quad &\lambda=1/16.\\ 
\end{array}
\]

The general solutions in the analytic form are known only 
in the integrable cases~\cite{Vernov:Conte4,Vernov:ConteTMF}, in other cases
not only four-, but even three-parameter exact solutions have
yet to be found. In nonintegrable cases local four-parameter solutions 
as converging psi-series solutions
have been found~\cite{Vernov:Melkonian} for all values of the parameter
$C$, except $C=-2$. The Ablowitz--Ramani--Segur algorithm of the  
Painlev\'e test appears very useful to find such values of parameter at 
which three-parameter solutions 
can be expanded in formal Laurent series 
and to construct these local solutions. The knowledge of local solutions 
can be used to find solutions in the analytical~form. 

Let us assume that the behavior of solutions in a sufficiently small
neighborhood of the singularity point is algebraic and 
$x$ and $y$ tend to infinity as 
\[
x=a_{\alpha}(t-t_0)^\alpha\qquad \mbox{and}\qquad
y=b_{\beta}(t-t_0)^\beta,
\]
where $\alpha$, $\beta$, $a_{\alpha}$ and $b_{\beta}$ are some constants.
 
The Painlev\'e test gives all information about
behavior of solutions in the neighborhood of the singularity point (see, 
for example, \cite{Vernov:Tabor}). There exist two possible variants of 
dominant behavior and resonance structure of solutions of the 
generalized H\'enon--Heiles system~\cite{Vernov:Tabor,Vernov:Melkonian}:

\noindent\begin{tabular}{|l|l|}
\hline
{ \it Case 1}:  & { \it Case  2}: $\!$($\beta<\Re e(\alpha)$)\\[2mm]
\hline $\alpha=-2$,&
 $\alpha=\frac{1\pm\sqrt{1-48/C}^{\vphantom{7^4}}}{2}$,\\[1mm] $\beta=-2$,
& $\beta=-2$,\\[1mm] $a_{\alpha}=\pm 3\sqrt{2+C}$, &
$a_{\alpha}=c_1^{\vphantom{4}}$ {(an arbitrary number)},\\[1mm]
$b_{\beta}^{\vphantom{27}}=-3$,
& $b_{\beta}^{\vphantom{27}}=\frac{6}{C}$, \\[1mm]
$r=-1,\; 6,\; \frac{5}{2}-\frac{\sqrt{1-24(1+C)}}{2},\;
\frac{5}{2}+\frac{\sqrt{1-24(1+C)}}{2}$ & $r=-1,\; 0,\; 6,\;
\mp\sqrt{1-\frac{48}{C}}$\\[2.7mm] \hline
\end{tabular}

The values of $r$ denote resonances: $r=-1$ corresponds to
arbitrary parameter $t_0$; $r=0$ (in the {\it Case 2}) corresponds to
arbitrary parameter $c_1^{\vphantom{4}}$.
Other values of $r$ determine powers of $t$, to be exact,
$t^{\alpha+r}$ for $x$ and $t^{\beta+r}$ for $y$, at which new
arbitrary parameters can appear.

For integrability of system (\ref{Vernov:2}) all values of $r$ have to be
integer and all systems with zero determinants have to have solutions
at any values of included in them free parameters. It is possible
only in the integrable cases (i) --- (iii).

Those values of $C$, at which $r$
are integer numbers either only in the {\it  Case 1} or only
in the {\it  Case 2}, are of interest for search of special 
three-parameter solutions. Those cases where an additional negative 
resonance is present likely correspond to singular, rather the general, 
solutions~\cite{Vernov:Tabor}. 

Let's consider all cases, when there exist special 
(no singular) solutions, representable as a three-parameter Laurent series 
(may be,
multiplied on $\sqrt{t-t_0}$). From the requirement that all values of
$r$ but one are integer and nonnegative numbers we obtain  
the following values of $C$: $C=-1$, $C=-4/3$ (the {\it Case 1}),
$C=-16/5$, $C=-6$, $C=-16$ (the {\it Case 2,
$\alpha=\frac {1-\sqrt {1-48/C}^{\vphantom{7^4}}}{2}$}) and
$C=-2$, when two types of singular behaviour
coincide.  

At $C=-2$ (in the {\it  Case 1}) \ $ a_{\alpha}=0$. This is
the consequence of the fact that, contrary to our assumption, the
behaviour of the solution in the neighborhood of a singularity point is not
algebraic, because its dominant term includes logarithm~\cite{Vernov:Tabor}.
 At $C=-6$ and any value of $\lambda $ the exact four-parameter solutions
are known. In cases $C = -1$ and $ C = -16 $ the substitution of unknown
function as Laurent series gives the equations in $\lambda$:
accordingly $\lambda = 1$ and $\lambda = 1/16$, hence,
in nonintegrable cases special three-parameter local solutions have to
include logarithmic terms.
Single-valued three-parameter special solutions
can exist only in two nonintegrable cases,  at $C = -16/5$ and
at $C = -4/3$.

\section{NEW SOLUTIONS}

Let us consider the  H\'enon--Heiles system
with $C=-16/5$.  In the {\it Case 1} some  values  of $r$ are
not rational.  To find special
three-parameter solutions we consider the {\it Case 2}.  In this case
$\alpha=-3/2$ and $r=-1,\;0,\;4,\;6$, hence, in the neighborhood
of the singularity point $t_0$ we have to seek $x$ 
in such a form that $x^2$
can be expanded into Laurent series, beginning with $(t-t_0)^{-3}$.
Let $t_0=0$, substituting
\begin{equation}
\label{Vernov:27a}
x=\sqrt{t}\left(c_1^{\vphantom{4}}t^{-2}+\sum_{j=-1}^\infty a_jt^j\right)
\qquad\mbox{and}\qquad y=-\:\frac{15}{8}t^{-2}+\sum_{j=-1}^\infty b_jt^j
\end{equation}
in (\ref{Vernov:2}), we obtain the following sequence of linear system in
 $a_k$ and $b_k$:
\begin{equation}
\label{Vernov:11}
\left\{
\begin{array}{l}
\displaystyle\Big
(k ^ 2-4\Big)a_{k} + 2c_1^{\vphantom{4}} b_{k} = -\lambda a_{k-2}
-2\sum_{j=-1}^{k-1}a_{j}b_{k-j-2},\\[7.2mm]
\displaystyle\Big((k-1)k-12\Big)b_{k}
= -\:b_{k-2} - \sum_{j=-2}^{k-1}a_{j}a_{k-j-3} -
\frac{16}{5}\sum_{j=-1}^{k-1} b_{j}b_{k-j-2}.
\end{array}
\right.
\end{equation}

 The determinants of the systems (\ref{Vernov:11}) corresponding
to $k=2$ and $k=4$ are equal to zero.
 To determine $a_2$ and $b_2$ we have the following system:
\begin{equation}
\label{Vernov:12}
\left\{
\begin{array}{l@{}}
\displaystyle c_1^{\vphantom{4}}\Big(557056 c_1^8 + \big(15552000\lambda -
4860000\big)c_1^4 + 864000000b_2 \: + \\[1mm]\displaystyle   +\:
108000000\lambda^2 - 67500000\lambda +10546875\Big)=0,\\[2mm]\displaystyle
 818176c_1^8  + \Big(15660000\lambda - 4893750\Big)c_1^4 -
 810000000b_2- 6328125=0.
\end{array}
\right.
\end{equation}

It is easy to see that this system contains no terms proportional to $a_2$,
therefore, $a_2$ is the new constant of integration.
We discard the solution with $c_1^{\vphantom{4}}=0$ and obtain
the system in $c_1^4$ and  $b_2$. System  (\ref{Vernov:12}) has solutions only if
\[
c_1^4=
\frac{1125\left(525- 1680\lambda\pm 4\sqrt{35(2048\lambda^2 - 
1280\lambda + 387)}\right)}{167552}.   
\]

We obtain new constant of integration $a_2$,
but we must fix $c_1^{\vphantom{4}}$, so number of
constants of integration is equal to 2.  It is easy to verify that
$b_4$ is an arbitrary parameter, because the  corresponding
system is equivalent to one linear equation.  
System (\ref{Vernov:2}) is invariant under exchange 
$x$ to $-x$, so, we obtain four different local 
solutions which depend on three parameters, namely  $t_0$,  $a_2$ and
$b_4$. With the help of some computer algebra system, for example,
{\bf REDUCE}~\cite{Vernov:REDUCE},
these solutions can be obtained with arbitrary accuracy.
For the case $\lambda=1/9$ the obtained Laurent series  are presented
in~\cite{Vernov:Vernov0209}.

At $C=-4/3$ the situation is similar.
In the {\it Case 1} we have $ r = -1, \; 1, \; 4, \; 6 $.
Substituting
\begin{equation}
\label{Vernov:27b}
x = \sqrt{6}t^{-2} + \sum_{k = -1}^\infty d_kt ^ k\qquad
\mbox{and} \qquad y = -3t ^ {-2} + \sum_{k = -1}^\infty f_kt^k
\end{equation}
in system (\ref{Vernov:2}), we receive a sequence of linear systems in $d_k$
and $f_k$:
\begin{equation}
\label{Vernov:14}
\left\{
\begin{array}{l}
\displaystyle
\Big((k-1)k-6\Big)d_{k} + 2\sqrt{6}f_{k}
= -\lambda d_{k-2} -2\sum_{j=-1}^{k-1}d_{j}f_{k-j-2},\\[7.2mm]
\displaystyle 2\sqrt{6}d_{k}+\Big((k-1)k-8\Big)f_{k}
= -\:f_{k-2} - \sum_{j=-1}^{k-1}d_{j}d_{k-j-2} -
\frac{4}{3}\sum_{j=-1}^{k-1} f_{j}f_{k-j-2}.
\end{array}
\right.
\end{equation}
The determinants of the systems (\ref{Vernov:14}) corresponding
to $k = -1,\: 2,\: 4$  are equal to zero.
The first system ($k=-1$) always has infinite number of
solutions and $f_{-1}$ is a parameter. We have to fix this parameter  to
solve the system corresponding to $k=2$. This system has solutions only if 
\[ 
f_{-1}^2=\frac { 105 - 140\lambda \pm \sqrt{7
(1216\lambda ^ 2-1824\lambda + 783)} }{385}
\qquad\mbox{or}\qquad f_{-1}= 0. 
\]

At $k=4$ system (\ref{Vernov:14}) is reduced to one equation.
Thus, at $ C = -4/3$ we have five three-parameter
($t_0$, $f_2$ and $f_4$) solutions. 

The convergence of all the Laurent series solutions 
on some real time interval have been proved in~\cite{Vernov:Melkonian}.
For the obtained solutions it is easy to find conditions, at which
the series converge at $0<|t|\leqslant 1-\varepsilon$,
where $\varepsilon$ is any positive number.
Our series converge in the above-mentioned ring, if $\exists N$ and $\exists M$
such that $\forall n> N$ \ $|a_n| \leqslant M$ and $|b_n| \leqslant M$.
 Let $|a_n|\leqslant M$ and $|b_n|\leqslant M$ for all $-1 < n < k$, then
 (in the case $C=-16/5$) from (\ref{Vernov:11}) we obtain:
 \[
 |a_k|\leqslant\frac{2M(k+1)+|\lambda| + 2|c_1^{\vphantom{4}}|}{|k^2-4|}M,
\qquad\qquad |b_k|\leqslant\frac{21Mk+26M+5}{5|k^2-k-12|}M. 
\]

It is easy to see that there exists such $N$ that if
$|a_n|\leqslant M$ and $|b_n|\leqslant M$
for $-1\leqslant n\leqslant N$, then $|a_n|\leqslant M$
and $ |b_n| \leqslant M$ for  $-1\leqslant n < \infty$.
So one can prove the convergence, analyzing values of a finite number of
the first coefficients of series. For   $C=-4/3$ it is easy to obtain
the analogous result.

\section{Global single-valued solutions}
We have found some local three-parameter solutions. 
To seek the global single-valued solutions we transform system 
(\ref{Vernov:2}) into the equivalent fourth order 
equation~\cite{Vernov:RM1992,Vernov:Timosh}:  
\begin{equation} 
\label{Vernov:3}
y_{tttt}=(2C-8)y_{tt}y - 
(4\lambda+1)y_{tt}+2(C+1)y_{t}^2+\frac{20C}{3}y^3+
(4C\lambda-6)y^2-4\lambda y-4H,  
\end{equation}
where $H$ is the energy of the system. We note, that the value of $H$
depends on initial data, that is to say that from five parameters:
$y(t_0)$, $y_t(t_0)$, $y_{tt}(t_0)$, $y_{ttt}(t_0)$ and $H$ only four are 
independent.

There are some reasons to seek three-parameter 
solutions of eq.~(\ref{Vernov:3}) in terms of elliptic functions. 
In 1999 E.I. Timoshkova~\cite{Vernov:Timosh} found that the general 
solution of the following equation:  
\[ y_t^2=\tilde {\cal  A}y^3+\tilde 
{\cal B} y^2+\tilde {\cal C} y+ \tilde {\cal D}+\tilde {\cal G} 
y^{5/2}+\tilde  {\cal E} y^{3/2} 
\] 
with some values of constants  $\tilde 
{\cal  A}$,  $\tilde {\cal B}$, $\tilde {\cal C}$, $\tilde {\cal D}$, 
$\tilde {\cal G}$ and $\tilde {\cal E}$, is one-parameter  solution of 
eq.~(\ref{Vernov:3}) in each of two above-mentioned nonintegrable cases 
($C= -4/3$ \ or \ $C= -16/5$, \ $\lambda$ is an arbitrary number). 
If  $\tilde {\cal G}= 0$ and $\tilde {\cal E}=0$  we obtain the
well-known solutions in terms of the Weierstrass elliptic function.
Solutions with $\tilde {\cal G}\neq 0$ or $\tilde {\cal E} \neq 0 $ are 
derived only at $\tilde {\cal D}=0$, therefore, substitution 
$y(t)=\varrho(t)^2$ 
gives:  \begin{equation} \label{Vernov:6} 
\varrho_t^2=\frac{1}{4}\Bigl(\tilde {\cal A}\varrho^4+\tilde {\cal G} 
\varrho^3+\tilde {\cal B} \varrho^2 + \tilde {\cal E}\varrho+\tilde {\cal 
C}\Bigr).  
\end{equation}

 Two-parameter solutions $y(t)=\varrho(t)^2+P_0$, where 
$P_0$ is an arbitrary constant and $\varrho(t)$ satisfy equation  
(\ref{Vernov:6}), have been obtained 
in~\cite{Vernov:Timoshkova2001,Vernov:Vetish}. 
 These solutions are the 
following elliptic functions 
\begin{equation}
\label{Vernov:7}
y(t-t_0) = \left (\frac {a\wp (t-t_0) + b} 
{c\wp (t-t_0) + d} \right)^2+P_0,  \qquad\qquad  (ad-bc=1),
\end{equation} 
where $\wp(t-t_0)$ is the Weierstrass elliptic 
function, $a$, $b$, $c$ and $d$ are some constants. The parameter $P_0$ 
defines the energy of the system.  There exist two different elliptic 
solutions for each possible pair of values of $C$ and $\lambda$.

 Let us consider the three-parameter solutions which
were obtained at $C=-4/3$. 
If we choose $f_{-1}=0$, then we 
obtain the solution 
which generalizes the known two-parameter solution in terms of Weierstrass 
elliptic functions.  Other solutions generalize  
two-parameter solutions, obtained in~\cite{Vernov:Vetish}.  The 
coefficient $f_{-1}$ is a residue of $y$.  The sum of residues 
of elliptic function in its parallelogram of periods has 
to be zero~\cite{Vernov:BE}, hence, 
 two local solutions with opposite signs of $f_{-1}$ correspond to one 
global elliptic solution.  The obtained local three-parameter solutions 
generalize the Laurent series of the two-parameter elliptic solutions
in the form (\ref{Vernov:7}). 

For $C=-16/5$ we obtained four local solutions, which generalize
two global elliptic solutions in the form (\ref{Vernov:7}). 
So, each obtained local three-parameter 
solution generalize the Laurent series of some two-parameter elliptic 
solution and we can assume that an unknown global there-parameter 
solutions are elliptic functions. 

 Of course, solutions, which are single-valued 
in the  neighborhood of one singularity point, can  
be multivalued in the  neighborhood of another  singularity point. So, we 
can only assume that global three-parameter solutions are single-valued. 
If we assume this and moreover that these solutions are elliptic functions 
(or some degenerations of them), then we can seek them as solutions of 
some polynomial first order equations. The classical theorem, which was 
established by Briot and Bouquet~\cite{Vernov:BriBo},  proves that if the 
general solution of the autonomous polynomial ODE is single-valued, then  this 
solution is either an elliptic function,  or a rational function of 
$e^{\gamma x}$, $\gamma$ being some constant,  or a rational function of 
$x$. Note that the third case is a degeneracy of the second one,  which in 
its turn is a degeneracy of the first one. At the same time, there exist 
elementary functions, for example, the function $f(t)=t+\sin(t)$, which 
are not solutions of any first order polynomial ODE.

A second result, of immediate practical use, is due to 
P.~Painlev\'e~\cite{Vernov:Painleve1}.  He has proved that if the general 
solution of an autonomous polynomial ODE is single-valued, then the 
necessary form of this ODE is 
\begin{equation} 
\label{Vernov:16} 
\sum_{k=0}^{m} \sum_{j=0}^{2m-2k}h_{jk}^{\vphantom{y}}\: 
y^j_{\phantom{y}} {y'}^k_{\vphantom{y}}=0,  
\end{equation} 
in which  $m$ is a positive integer number and  $h_{jk}$ are constants. 

In 2003 R. Conte and M. Musette have proposed a new method 
to find elliptic solutions~\cite{Vernov:CoMu03}. This method is based on the
Painlev\'e test and 
uses the Laurent series expansion to find the
analytic form of elliptic solutions. Rathen than to substitute  
the first order equation (\ref{Vernov:16}) into equation (\ref{Vernov:3}) 
one can substitute the found Laurent series solutions of equation
(\ref{Vernov:3}), for example, either solution (\ref{Vernov:27a}) or 
solution (\ref{Vernov:27b}), into equation (\ref{Vernov:16}) and obtain 
a linear system in  $h_{jk}$. This method is more powerful than the 
traditional method and allows in principal to find all elliptic solutions. 
I hope that the use of this method allows to find the three-parameter 
elliptic solutions.

\section{CONCLUSION}

Using the Painlev\'e analysis  we have found local
special solutions in two nonintegrable cases of the generalized
H\'enon--Heiles system ($C=-16/5$ and $C=-4/3$). These solutions are
the converging Laurent series,
depending on three parameters. For some values of these parameters the
obtained solutions coincide with the known exact periodic solutions.
There are no obstacles to exist three-parameter single-valued 
solutions, so, the probability of finding of exact
three-parameter solutions, which  generalize the solutions
obtained in~\cite{Vernov:Timosh,Vernov:Vetish}, is high.

The author is grateful to \ F.~Calogero, \  R.~Conte, \ V.~F.~Edneral, \  
A.~K.~Pogrebkov  \ and \ E.~I.~Timoshkova for valuable discussions.  This 
work has been supported by the Russian Foundation for Basic Research under 
grants 00-15-96560,  00-15-96577  and  NSh--1685.2003.2 and by the grant 
of  the scientific Program "Universities of Russia" \ UR.02.03.002.

\LastPageEnding

\begin{thebibliography}{99}
\footnotesize
\bibitem{Vernov:Kova} {\it Kowalevski S.}, Sur le
probl\`eme de la rotation d'un corps solide autour d'un point fixe, {\it
Acta  Mathematica}, 1889, V.~12,  177--232; \  Sur une properi\'et\'e
du sust\`eme d'\'equations diff\'erentielles qui d\'efinit la rotation
d'un corps solide autour d'un point fixe, {\it Acta  Mathematica}, 1890,
V.~14,  81--93, \{in French\}. \  Reprinted in: {\it
Kovalevskaya S.V., } Scientific Works, AS USSR Publ.
House, Moscow, 1948, \{in~Russian\}.
\bibitem{Vernov:Golubev2} {\it Golubev V.V., } Lectures on the Integration
of the Equation of Motion of a
Heavy Rigid Body about a Fixed Point, Gostekhizdat,
Moscow, 1953, reprinted: RCD, Moscow--Izhevsk,
2002, \{in~Russian\}.
\bibitem{Vernov:Goriely0}  {\it Goriely A., } A brief history of
Kovalevskaya exponents and modern developments,
{\it Regular and Chaotic Dynamics}, 2000, V.~5, N~1,
3--16.
\bibitem{Vernov:Painleve1} {\it  Painlev\'e P., }
Le\c{c}ons sur la th\'eorie analytique
des \'equations diff\'e\-rentielles, profees\'ees \`a Stockholm
(septembre, octobre, novembre 1895) sur l'invitation de
S. M. le roi de Su\`ede et de Norw\`ege, Hermann, Paris, 1897. \
Reprinted in:  O$\!$euvres\ de Paul Painlev\'e, V.~1, ed. du
CNRS, Paris, 1973.
    On-line version: The Cornell Library Historical
    Mathematics Monographs, http://historical.library.cornell.edu/
\bibitem{Vernov:Golubev1}  {\it  Golubev V.V., } Lectures
on Analytical Theory of Differential Equations, Gostekhizdat, 
Moscow--Leningrad, 1950, \{in~Russian\}.
\bibitem{Vernov:Hille} {\it
Hille E., } Ordinary Differential Equations
in the Complex Domain, New York,
Wiley, 1976.
\bibitem{Vernov:Tabor}{\it  Tabor M., }
Chaos and Integrability in Nonlinear
Dynamics, Wiles, New York, 1989, \{in~English\}, URSS, Moscow, 2001,
 \{in~Russian\}.
\bibitem{Vernov:ES1} {\it
Ercolani  N. and  Siggia E.D., } Painlev\'e Property and Integrability,  {\it
Phys. Lett. A}, 1986, V.~119,  112--116.  
 \bibitem{Vernov:ES2} {\it  Ercolani N. and  Siggia E.D., } Painlev\'e 
 Property and Geometry, {\it Physica D}, 1989, V.~34,  303--346.
\bibitem{Vernov:Kozlov} {\it  Kozlov V.V., }
Symmetry, topology and resonances in
Hamiltonian mechanics, Izhevsk, UGU Publ.  House, 1995, \{in~Russian\},
 Berlin, Springer, 1995, \{in~English\}.
\bibitem{Vernov:GoTa} {\it  Goriely A. and Tabor M., } The Role of Complex-time
Singularities in Chaotic Dynamics, {\it Regular Chaotic Dynamics}, 1998,
V.~3, N~3, 32--44.
\bibitem{Vernov:Goriely} {\it Goriely A., } Painlev\'e Analysis and
Normal Forms Theory, {\it Physica D}, 2001, V.~152--153, 124--144.
\bibitem{Vernov:Conte0}{\it Conte R., } (ed.) The Painlev\'e
property, one century later,
Proceedings of the Carg\`ese school (3--22 June, 1996, Carg\`ese), CRM series
in mathematical physics, Springer--Verlag, 
Berlin, 1998,  New York, 1999.
\bibitem{Vernov:ARS} {\it  Ablowitz M.J.,  Ramani A. and  Segur H., } 
A Connection between Nonlinear Evolution Equations and Ordinary 
Differential Equations of P-type. I $\&$ II, {\it J.  Math.  Phys.}, 1980, 
V.~21,  715--721 $ \& $ 1006--1015.  
\bibitem{Vernov:WTC}{\it Weiss J., 
 Tabor M. and Carnevale G., } The Painlev\'e Property for Partial 
 Differential Equations, {\it J.  Math.  Phys.}, 1983, V.~24, 
 522--526.  
\bibitem{Vernov:Weiss4}{\it  Weiss J., } The Painlev\'e 
Property for Partial Differential Equations. II:  B\"acklund 
Transformation, Lax Pairs and the Schwarzian derivative, {\it J.  Math.  
Phys.}, 1983, V.~24,  1405--1413.  
\bibitem{Vernov:Conte2}{\it Conte R., Fordy A.P. and Pickering A., } A 
 Perturbative Painlev\'e Approach to 
Nonlinear Differential Equations, {\it Physica D}, 1993, V.~69, 
 33--58.  
\bibitem{Vernov:Conte1988}{\it Conte R., } Universal 
invariance properties of Painlev\'e analysis and B\"acklund transformation 
in nonlinear partial differential equations, {\it Phys.~Lett.~A}, 1988,  
V.~134,  100--104.  
\bibitem{Vernov:Tabor2}  {\it Cariello F. and Tabor M., } 
 Painlev\'e Expansions for Nonintegrable Evolution Equations, {\it   
Physica D}, 1989, V.~39,   77--94.  
 \bibitem{Vernov:Conte3} {\it Conte R., } Exact solutions of
   nonlinear partial differential equations by 
singularity analysis, nlin.SI/0009024.  
\bibitem{Vernov:Weiss1}{\it  Weiss J., }  B\"acklund Transformation
and Linearizations of the H\'enon--Heiles System,
{\it Phys. Lett. A}, 1984, V.~102,  329--331.
\bibitem{Vernov:Weiss2}{\it  Weiss J., }
B\"acklund Transformation and the H\'enon--Heiles System,
{\it Phys. Lett. A}, 1984, V.~105,  387--389.
\bibitem{Vernov:GNTZ} {\it Gibbon J.D., Newell A.C., Tabor M. and 
Zeng Y.B., } Lax Pairs,  B\"acklund Transformation and Special Solutions 
for Ordinary Differential  Equations, {\it Nonlinearity}, 1988, V.~1, 
481--490.  
\bibitem{Vernov:Sahadevan}{\it  Sahadevan R., }  Painlev\'e 
Expansion and Exact Solution for Nonintegrable Evolution Equations, { \it
TMF (Russ. J. Theor.  Math. Phys.)}, 1994, V.~99,   528--536,
Theor. Math. Phys., 1994, V.~99, 796--802.
\bibitem{Vernov:Bianchi3} {\it Springael J., Conte R. and
Musette M., } On
the exact solutions of the Bianchi IX cosmological model in the proper
time, {\it Regular Chaotic Dynamics},  
1998, V.~3, N.~1,  3--8; nlin.SI/9804008.
\bibitem{Vernov:HeHe}  {\it
 H\'enon M. and Heiles C., } The Applicability of the Third Integral of 
Motion:  Some Numerical Experiments, {\it Astronomical J.}, 1964, V.~69, 
 73--79.
\bibitem{Vernov:Vernov0203} {\it Vernov S.Yu., } The Painlev\'e
Analysis and Special Solutions for Nonintegrable Systems, math-ph/0203003. 
\bibitem{Vernov:Podolski}{\it   Podolsk\'y Ji. and Vesel\'y K., }
Chaos in $pp-$wave spacetime, {\it Phys.  Rev. D}, 1998, V.~58, 081501.
\bibitem{Vernov:Kokubun1} {\it Kokubun F., }  Gravitational waves from the
H\'enon--Heiles system,  {\it  Phys. Rev.  D}, 1998, V.~57,
 2610--2612.
\bibitem{Vernov:Polska} {\it Antonowicz  M. and Rauch-Wojciechowski S., }
Bi-Hamiltonian formulation of the H\'enon--Heiles system and its
multidimensional extensions, {\it Phys. Lett. A}, 1992, V.~163,
 167--172.
\bibitem{Vernov:Conte4}
 {\it Conte  R., Musette M. and Verhoeven C., }  Integration of a generalized
H\'enon--Heiles  Hamiltonian,
{\it J.  Math.  Phys.}, 2002,  V.~43,  1906--1915; nlin.SI/0112030. 
 \bibitem{Vernov:ConteTMF}
{\it Conte R., Musette M. and Verhoeven C., } 
General solution for Hamiltonians with
extended cubic and quadratic potentials,
{\it TMF (Russ. J. Theor.  Math. Phys.)}, 2003, V.~134, 148--159,
\{in~Russian\}, 128--138, \{in~English\}; nlin.SI/0301011.
\bibitem{Vernov:Melkonian} {\it Melkonian S., } Psi-series
Solutions of the Cubic H\'enon--Heiles System and Their  Convergence,
{\it  \ J. of Nonlin. Math.  Phys.},  1999, V.~6,  139--160; math.DS/9904186.
\bibitem{Vernov:REDUCE} {\it Hearn A.C., }
REDUCE. User's Manual.  Vers. 3.6, RAND Publ. CP78,
Santa Monica, California,  1995,
http://www.unikoeln.de/REDUCE/3.6/doc/reduce/
 \\  
REDUCE. User's and Contributed Packages Manual, Vers. 3.7,
CA and Codemist Ltd, Santa Monica, California, 1999, \ \
http://www.zib.de/Symbolik/reduce/more/moredocs/reduce.pdf
\bibitem{Vernov:Vernov0209} {\it  Vernov S.Yu., } Construction
of solutions for the generalized H\'enon--Heiles system with the help of
the Painlev\'e test, 
{\it TMF (Russ. J. Theor.  Math. Phys.)}, 2003, V.~135,  409--419, 
 \{in~Russian\}, 792--801, \{in~English\};  math-ph/0209063.
\bibitem{Vernov:RM1992}{\it Conte R. and Musette M., }
Link between solitary waves and projective Riccati equations,
J.~Phys.~A, 1992, V.~25,   5609--5623.
\bibitem{Vernov:Timosh}  {\it Timoshkova E.I., } A New Class of
Trajectories of Motion in the H\'enon--Heiles Potential Field, {\it Rus.
Astron. J.}, 1999, V.~76,  470--475, \{in~Russian\}, {\it
Astron. Rep.}, 1999, V.~43,  406--411, \{in~English\}.
\bibitem{Vernov:Timoshkova2001}  {\it Timoshkova E.I., } 
A New Class of Periodic
Orbits in the H\'enon-Heiles Potential Field, in  Proceedings of 
the International Conference "Stellar Dynamics: from Classic
 to Modern" (21--27 August, 2000, Sankt--Petersburg, Russia), editors 
L.P.~Ossipkov and I.I.~Nikiforov,  Sankt--Petersburg, 2001, 201--205.  
 \bibitem{Vernov:Vetish}  {\it Vernov  
S.Yu. and Timoshkova E.I., } New two--parameter solutions for the generalized 
 H\'enon--Heiles system, Preprint SINP MSU 2003--14/727, \\
 \verb#http://www.npi.msu.su/science.php3?sec=preprints&year=2003#, 
  \ \{in~Russian\}.  
\bibitem{Vernov:BE}{\it Erdelyi A.~et al., }  (eds.)  Higher
 Transcendental Functions (based, in part, on notes left by H.~Bateman).  V.~3, MC
Graw-Hill Book Comp., New York, London, 1955,
 \{in~English\}, "Nauka", Moscow, 1967, \{in~Russian\}.
\bibitem{Vernov:BriBo}  {\it Briot C. and Bouquet T., } Th\'eorie des fonctions 
elliptiques. L. I-VIII, Gauthier-Villars, Paris, 1875.
\bibitem{Vernov:CoMu03} {\it Conte R. and Musette M., }
Analytic solitary waves of nonintegrable equations,  Physica D, 2003,
V.~181, 70--79, nlin.PS/0302051.
\end{thebibliography}
\end{document}